\documentstyle[epsfig]{aipproc}

\begin{document}

\title
{ Flash of Prompt Photons \\ from \\
the Early Stage of Heavy-Ion Collisions}

\author{
 Dinesh Kumar Srivastava\thanks{Contribution to Proceedings of
``Kay Kay Gee Day- Klaus Kinder-Geiger Memorial Workshop''
held at Brookhaven National Lab, October 23, 1998}}

\address{
 Variable Energy Cyclotron Centre\\ 1/AF Bidhan Nagar\\
 Calcutta 700 064\\ India}

\maketitle

\begin{abstract}
We briefly recall the tremendous strides in the studies of the
parton cascade model made by Klaus Geiger.
Next, we argue that photons may provide confirmation of several of
these ideas. Thus we know that,
copious internetted partonic cascades may develop in the wake of
relativistic collisions of nuclei at CERN SPS  and BNL RHIC energies, 
We use  the parton cascade model to estimate the emission of single 
photons generated from such cascades due to the branching of quarks
$ q\,\rightarrow\,q\,\gamma $, scattering of quarks and gluons, and
annihilation of quarks. The formation of a hot and dense
partonic matter is shown
to be preceded by an intense radiation of photons from the QED
branching of quarks. This is similar to the
QCD branching $ q\rightarrow\,qg $ which along with the gluon
multiplication (  $ g\rightarrow\,gg $ ) 
which is known to play a crucial role in the
formation of the dense partonic plasma.
\end{abstract}

\section*{From PCM to VNI; An Ode to KKG}
It is incumbent on us to provide a comprehensive and accurate
description of relativistic collision of nuclei from the instant
of nuclear contact to the formation of hadronic states which are
ultimately detected in experiments. Such collisions are expected to
create: a) the conditions which prevailed at the
time of early universe- a few micro-seconds after the big-bang,
and thus, b)  a strongly interacting matter under conditions of
extreme temperatures
 and densities  which would throw light on the ``hallowed''
quark-hadron phase transition.

A significant step in this direction was provided by the parton
cascade model (PCM)~\cite{pcm1} which was proposed to study the time
evolution of the parton phase space distribution in relativistic
nuclear collisions. In this approach the space-time description
is formulated within renormalization group-improved QCD perturbation
theory embedded in the framework of relativistic transport theory.
The dynamics of the dissipative processes during the early stage
of the nuclear reactions is thus simulated as the evolution of
multiple internetted parton cascades associated with quark and
gluon interactions. The model was considerably improved and
extended~\cite{pcm2} to include a number of new effects
like individual time scale of each parton-parton collision,
formation time of parton radiation, effective suppression of
radiative emissions from virtual partons due to enhanced
absorption probability of others in regions of dense
phase space occupation and the effects of soft gluon interference
for low energy gluon emissions, which all become important
in nuclear collisions. With these improvements the model
was used~\cite{pcm3} to study the dynamics of partons in relativistic
collision of gold nuclei at BNL RHIC and CERN LHC energies.
In particular very useful information about the evolution of
partons from pre-equilibrium to a thermalized quark-gluon plasma
was obtained along with the temperature, energy-density, and
entropy-density etc. It was demonstrated that energy densities in
excess of an order of magnitude of the critical energy-density
when a quark-hadron phase transition is expected, could be attained
in such collisions. The model was further used
to study the evolution of chemical evolution~\cite{pcm4},
the production of strangeness, charm, and bottom~\cite{pcm5},
 dileptons~\cite{pcm6},
and  very recently- single photons~\cite{pcm7} in such collisions.

The next important step~\cite{pcm8} involved
combining the above parton cascade model with a phenomenological
cluster hadronization model~\cite{chm1,chm2,chm3} which is motivated
by the ``preconfinement'' property~\cite{prec} of partons which is seen
from the tendency of quarks and gluons produced in parton cascades
to arrange themselves in colour neutral clusters, already at the
perturbative level~\cite{cnc}. This approach provided a decent
description of the experimentally measured momentum and multiplicity
distributions for $ p\overline{p} $ collisions at $ \sqrt{s} $ 
= 200 -- 1800 GeV,
and was further used to predict the multiplicity distributions likely to
be attained in relativistic heavy ion collisions at BNL RHIC and CERN LHC.
A critical review of these developments along with details can be
found in~\cite{pcm9}.

However, the above description of hadron formation~\cite{pcm8} did not
{\em explicitly} account for the colour degree of freedom of the partons,
which is, after-all, at the origin of confinement. To be specific, the
``ansatz'' for the confinement picture  was based exclusively on the
dynamically evolving space-time separations of nearest-neighbour
colour charges in the parton cascade, rather than on the details of the
colour structure of produced gluons, quarks, and anti-quarks. Thus,
it was assumed that due to the above mentioned ``pre-confinement''
property of QCD, the partons which are close in colour (in particular
minimal colour singlets) are also close in phase space. In other words,
instead of using a colour flow  description, the colour structure
during the development of the cascade was ignored and at the end
of perturbative evolution, colour neutral clusters were formed
from partons which had a minimal separation in coordinate and momentum
space.

It is known that this correspondence between the colour and
space-time structures of a parton cascade is not an equivalence,
but holds only in the average~\cite{color}. Thus, it has been argued
that the colour structure of the cascade tree provides in principle,
exact microscopic information about the flow of colour charges, whereas
the space-time structure is based on our model for the statistical
kinetic description of parton emission and the nearest-neighbour search,
which may be subject to fluctuations that deviate from the exact
colour-flow~\cite{pcm10}. This issue is expected to become increasingly
important when more particles populate a phase-space region, e.g.
for small $ x $ region in deep inelastic collisions and in hadron-nucleus
and nucleus-nucleus collisions. Thus it is expected that in such
cases, it is increasingly likely that the nearest neighbours in
momentum and phase space  would not necessarily form a colour singlet.
It is also likely that the ``natural'' colour-singlet partner for a
given parton within the same cascade (its ``endogamous'' partner)
might actually be disfavoured in comparison with a colour singlet
partner from a different but overlapping cascade (an ``exogamous''
partner). These consideration were incorporated in the
hadronization scheme with colour flow discussed in Ref.~\cite{pcm10},
where it was provided that if the space-time separation of two
nearest-neighbour partons allows coalescence, they can always
produce one or two color-singlet clusters, accompanied, if necessary,
by the emission of a gluon or a quark that carries away any unbalanced
net colour (see fig.5, Ref.~\cite{pcm10}).

This parton cascade -cluster hadronization model with colour flow
is now available in the form of a fortran programme- VNI~\cite{vni},
and has already formed basis for some interesting studies at SPS
energies~\cite{pcm11}, and also to investigate the effect of the
hadronic cascades at the end of the hadronization~\cite{pcm12}.

There are very few examples in recent times, where one person has contributed
so much in such a short time.

\section*{Photons from cascading partons}
Photons, either radiated or scattered, have remained one of the most
effective probes of every kind of terrestrial or celestial matter over the
ages. Thus, it is only befitting that the speculation of the
formation of  deconfined strongly interacting matter - some form of
the notorious quark-gluon plasma (QGP) - in relativistic heavy ion collisions,
was soon followed by a suggestion \cite{shuryak} that it should
be accompanied by   a characteristic radiation of photons. 
The effectiveness of photons in probing the history of such 
a hot and dense matter stems from the fact that, after production, 
they leave the system without any further interaction and thus carry 
unscathed information about the circumstances of their birth.
This is a very important consideration indeed, as the formation of 
a QGP is likely to proceed from a hard-scattering of initial
partons, through a pre-equilibrium stage, to perhaps a thermally and 
chemically equilibrated state of hot and dense partonic matter.
This matter will hadronize and interaction among hadrons will also give
rise to photons.  In this letter we concentrate on photons coming from the
early partonic stage in such collisions.
\smallskip

During the  partonic stage, photons emerge from two different mechanisms:
firstly, from collisions between  partons, 
i.e., Compton scattering of quarks and gluons
and annihilation of quarks and antiquarks; secondly,
from radiation of excited partons, i.e.
electromagnetic brems-strahlung of time-like cascading partons.
Whereas the former mechanism has been studied in
 various contexts \cite{joe,crs},
the latter source of photons is less explored~\cite{sjoestrand}, 
although, as we shall show, it is potentially much richer both in magnitude and complexity.

\smallskip

The {\it Parton Cascade Model} (PCM)~\cite{pcm9}, provides a
fully dynamical description of  relativistic heavy ion collisions.
It is based on the parton picture of hadronic
interactions and describes the nuclear dynamics in terms of
the interaction of quarks and gluons within the perturbative quantum
chromodynamics, embedded in the framework of relativistic
transport theory. The time evolution of the system is simulated by
solving an appropriate transport equation in a six-dimensional
phase-pace using Monte Carlo methods.
The procedure implemented in the computer code VNI~\cite{vni}
follows the dynamic evolution of
of scattering, radiating, fusing, and clusterizing partons till they are
all converted into hadrons. 
VNI, the Monte Carlo implementation of the PCM, has been adjusted on the
basis of experimental data from $ e^{+} e^-$ annihilation and $pp$ 
$ (p\bar{p}) $ collisions.

As recounted earlier, the PCM has been extensively used to provide
 valuable insight into
conditions likely to be achieved at RHIC and LHC energies~\cite{pcm9}.
Very recently, it has been found~\cite{pcm11} to provide reasonable
 description to
a large body of particle spectra from $ Pb+Pb $ and $ S+S $ collisions at
CERN-SPS energies as well.

Prompt photons are ideally suited to test the evolution of the partonic
matter as described by the PCM. They would accompany the early hard scatterings
and the approach to the thermal and chemical equilibration. Most importantly,
the PCM is free of assumptions of any type about the
initial conditions, since the space-time evolution of the matter
is calculated causally from the moment of collision onwards and at 
any point the state of the matter is determined by the preceding
space-time history.

\section*{$ q\,\rightarrow\,q\gamma $ and $ q\,\rightarrow\,qg $}

There are some important and interesting differences between
a scattering or a branching leading to production
of photons and gluons, as has been pointed out nicely by 
Sj\"ostrand~\cite{sjoestrand}. 
\begin{description}
\item[(i)]
Consider an energetic quark produced in a hard scattering.
It will radiate gluons and photons till its virtuality
drops to some cut-off value $ \mu_0 $.  The branchings 
$ q\,\rightarrow\,q\gamma $
and $q\,\rightarrow\,qg $ appear in the PCM on an equal footing and as
competing processes with similar structures. The probability, for a quark
to branch at some given virtuality scale $ Q^2 $, with the daughter quark
retaining a fraction $ z $ of the energy of the mother quark, is given by: 
\begin{equation}
d{\cal{P}}=\left( \frac{\alpha_s}{2\pi}C_F +\frac{\alpha_{\rm {em}}}{2\pi}e_q^2
         \right )  \frac{dQ^2}{Q^2} \frac{1+z^2}{1-z} dz
\end{equation}
where the first term corresponds to gluon emission and the second to photon
emission. Thus, the relative probability for the two processes is,
\begin{equation}
\frac{ {\cal {P}}_{q\rightarrow q\gamma} }
     { {\cal {P}}_{q\rightarrow q g} }
\;\propto\;\frac{\alpha_{\rm{em}} \langle e_q^2 \rangle }{\alpha_s C_F}
\;\simeq \;\frac{1}{200}
\;,
\end{equation}
for
$ \alpha_{\rm{em}}=1/137 $, $ \alpha_s = 0.25 $, 
$ \langle\,e_q^2\,\rangle$ = 0.22
and $ C_F $ = 4/3.  This does not mean, though, that
we can simulate emission of photons in a QCD shower by simply
replacing the strong coupling constant $ \alpha_s $ with the
electromagnetic $ \alpha_{\rm{em}} $ and the QCD colour Casimir 
factor $ C_F $ by $ e_q^2 $.  One has to  keep in mind that
the gluon,  thus emitted, may branch further, either as
$ g\,\rightarrow\,gg $, or as $ g\,\rightarrow\,q\bar{q} $;
implying that the emitted gluon has an effective non-zero mass.
As the corresponding probability for the photon to branch into a quark
or a lepton pair is very small, this process is neglected and we take the
photon to have a zero mass. (However, if we wish to study the dilepton
production from the collision, this may become an important
contribution~\cite{pcm6}; see later.)
\item[(ii)]
Secondly, the radiation of gluons
from the quarks is subject to soft-gluon interference which is enacted
by imposing an angular ordering of the emitted gluons. This is
not needed for the emitted photons. To recognize this aspect,
consider a quark which has `already' radiated a number of `hard' gluons.
The probability to radiate an additional `softer' gluon will get contributions
from each of the existing partons which may further branch  as 
$ q\,\rightarrow\,qg $ or $g\,\rightarrow\,gg $. It is well-known (see e.g.;
 Ref.~\cite{cnc})
that if such a soft gluon is radiated at a large angle with respect to all
the other partons and  one adds the individual contributions incoherently,
the emission rate would be overestimated, as the interference is
destructive. This happens as a soft gluon of a long wave-length is not able to
resolve the individual colour charges and sees only the net charge. 
The probabilistic picture of PCM is then recovered by demanding that
emissions are ordered in terms of decreasing opening angle 
between the two daughter partons at each branching, i.e., restricting the
phase-space allowed for the successive branchings. The photons, on the
other hand, do not carry any charge and only the quarks radiate.
Thus this angular ordering is not needed for them.
\item[(iii)]
Finally, the parton emission probabilities in the QCD showers contain
soft and collinear singularities,  which are regulated by introducing
a cut-off scale $ \mu_0 $. This regularization procedure
implies  effective masses for quarks and gluons,
\begin{equation}
m_{\rm{eff}}^{(q)}=\sqrt{\frac{\mu_0^2}{4} +m_q^2}
\;\;\;\;\;\;\;\;
m_{\rm{eff}}^{(g)}=\frac{\mu_0}{2}
\;,
\end{equation}
where $ m_q $ is the current quark mass. Thus the gluons cannot branch
unless their mass is more than $ 2m_{\rm{eff}}^{(g)}=\mu_0 $,
while  a quark cannot  branch unless its mass is more than
$ m_{\rm{eff}}^{(q)}+m_{\rm{eff}}^{(g)} $. 
An appropriate value for $ \mu_0 $ is about 1 GeV~\cite{pcm9}; a larger value is
not favoured by the data, and a smaller value will cause the perturbative
expression to blow up.  These arguments, however,  do not apply for photon
emission, since QED perturbation theory does not break-down and photons
are not affected by confinement forces. Thus, in principle quarks
can go on emitting photons till their mass reduces to current quark mass.
One may further argue that if the confinement
forces screen the ``bare'' quarks  the effective cut-off can be
of the order of a GeV. Thus
we can choose the cut-off scale $ \mu_0 $ separately for
the emission of photons and 
get valuable insight about confinement at work.
\end{description}

 The discussion above was focussed on the production of photons from the
branching of quarks. We also include the parton scattering processes in
the PCM which yield photons:
$ 
q+\bar{q}   \rightarrow  g +\gamma 
,\;
q+g \rightarrow  q + \gamma~,
$
from annihilation and Compton processes, the perturbative cross-sections
of which are well-known. In the PCM approach we treat these
processes within a perturbative QCD if the transverse momentum of
the process ($ p_T $) is lager than some cut-off $ p_T^0 $; see~\cite{p0}. 
Thus our results for these contributions will be strictly valid only
for $ p_T > p_T^0 $. (This cut-off is introduced 
in the collision frame of the partons, and thus we shall
have contributions even for smaller $ p_T $ in the nucleus-nucleus 
c.m. frame.)
\medskip

\section*{Results}
We study four examples: $S+S$ and $ Pb+Pb $ collisions at SPS
energies and $ S+S $ and $ Au+Au $ collisions at RHIC energies.

\begin{figure}[b!]
\centerline{ \epsfig{file=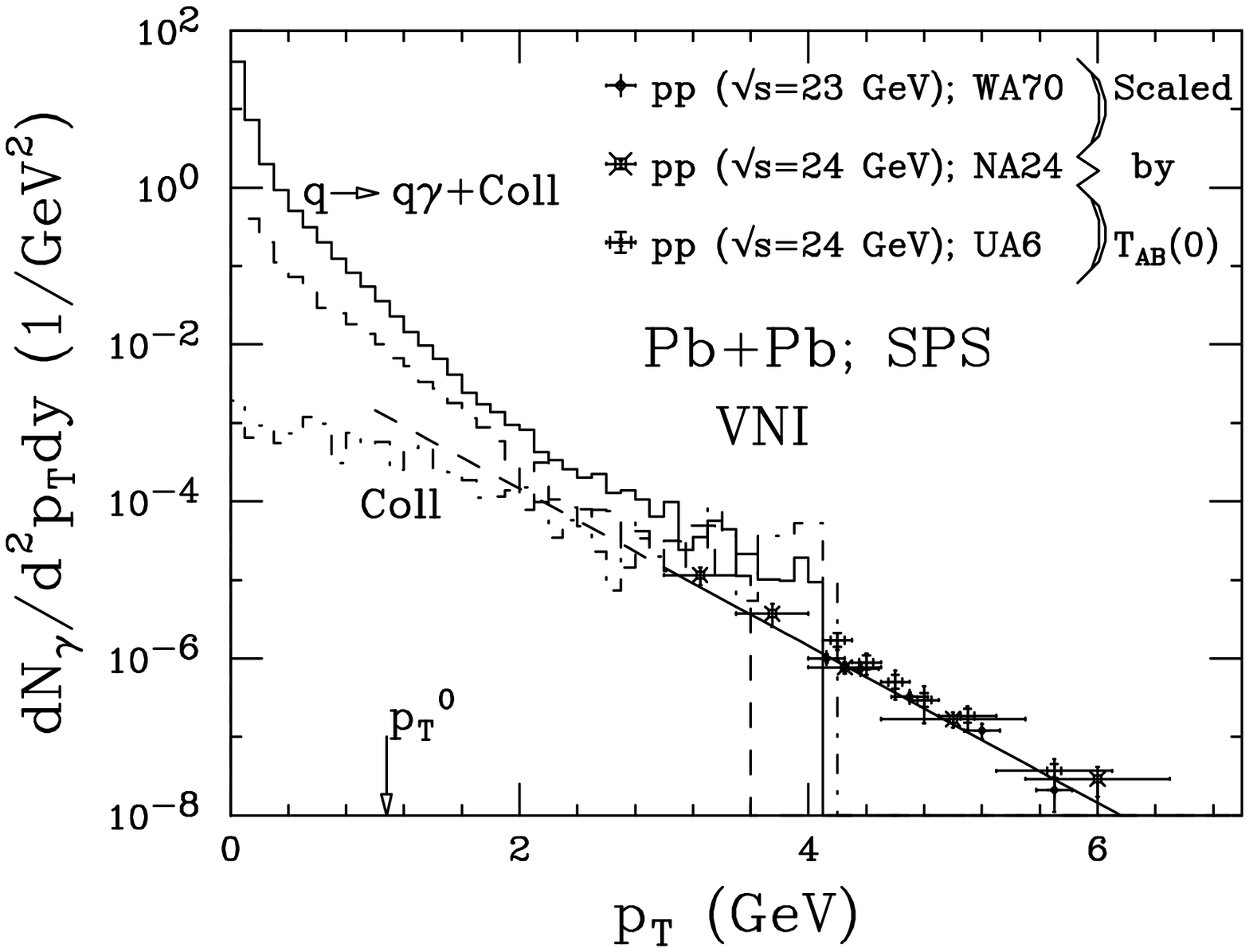,height=5.5in,width=5.5in}}
\vspace{10pt}
\begin{caption}
The radiation of prompt photons from
partonic matter in central collision of Pb (158  GeV/nucleon)+Pb 
nuclei at CERN SPS. The dot-dashed histogram gives the contribution
of only the collision processes.  The dashed and the solid histograms
give the contribution of the collision plus branchings when the
\protect{$\mu_0$} for the \protect{$q\rightarrow\,q\gamma$} branching
is taken as 1 and 0.01 GeV respectively.  The \protect{$\,p\,p\,$} data at
\protect{$\sqrt{s}\approx$} 24 GeV scaled by the
nuclear overlap function for the central collision of lead nuclei 
and the corresponding QCD prediction (solid curve, arbitrarily
extended to lower \protect{$p_T$}) is also shown for a comparison.
\protect{$p_T^0$} denotes the momentum cutoff above which 
the hard scatterings are included in the PCM.
\end{caption}
\label{fig1}
\end{figure}

In Fig.~1 we have plotted the production of single photons from such a
partonic matter in the central rapidity region for $ Pb+Pb $ system at SPS
energies.  The dot-dashed histogram shows the contribution of Compton and
annihilation processes mentioned above. The dashed and the solid
histograms show the total contributions (i.e., including the branchings
$ q\,\rightarrow\,q\gamma $) when the virtuality cut-offs for the photon
production is taken respectively as 0.01 and 1 GeV. 

We see that prompt photons from the quark branching completely 
dominate the yield for
$ p_T \leq 3 $ GeV, whereas at larger transverse momenta the
photons coming from the collision processes dominate.  The reduction
of the virtuality cut-off for the $ q\,\rightarrow\,q\gamma $ branching
is seen to enhance the production of photons having lower transverse
momenta as one  expects.

We have also shown the production of single photons from $ pp $ 
collisions for $ \sqrt{s}\approx $ 24  GeV,
obtained by WA70~\cite{wa70}, NA24~\cite{na24}, and UA6~\cite{ua6}
collaborations scaled by the nuclear thickness for zero impact parameter for
the collision of lead nuclei. The solid curve gives the
perturbative QCD results~\cite{jean} for the $ pp $ collisions
scaled similarly. The dashed curve is a direct extrapolation
of these results to lower $ p_T $.

\begin{figure}[b!]
\centerline{ \epsfig{file=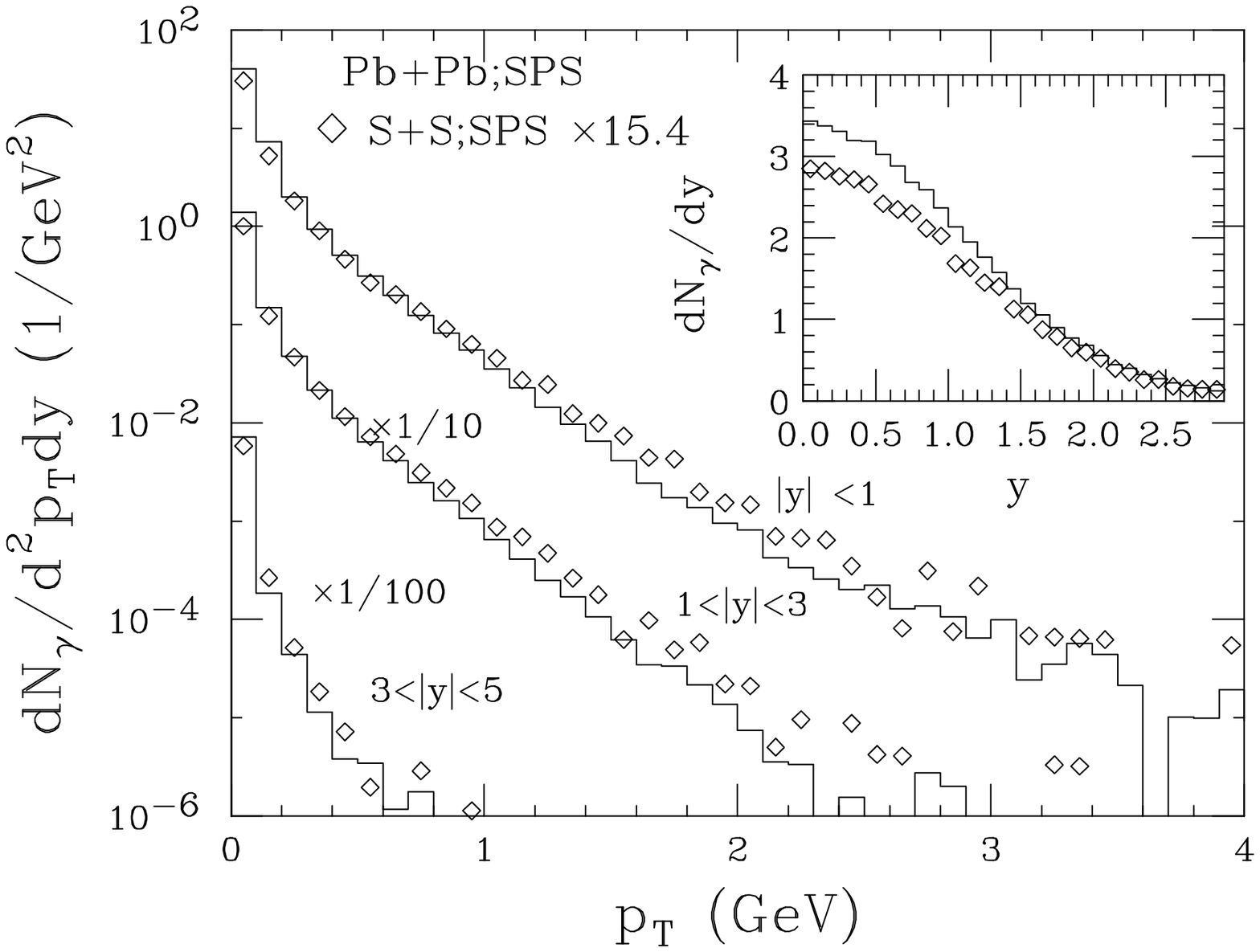,height=5.5in,width=5.5in}}
\vspace{10pt}
\begin{caption}
The radiation of single photons from S+S (200 GeV/Nucleon)
and Pb+Pb (158 GeV/Nucleon) collisions in different rapidity bins.
The inset shows the rapidity distribution of the radiated photons.
The results for the S+S collisions have been scaled by the
ratio \protect{$T_{Pb-Pb}/T_{S-S}\approx$} 15.4 for central collisions.
\protect{$\mu_0$} for the quark branching 
\protect{$ q\,\rightarrow\,q\gamma $} is taken as 0.01 GeV.
\end{caption}
\label{fig2}
\end{figure}

In Fig.~2 we have plotted the transverse momenta of the single photons
in several rapidity bins for $ Pb+Pb $ and $ S+S $ systems at SPS energies.
We see that the transverse spectra scale reasonably well with the
ratio of the nuclear overlap for central collisions for the
two systems $ T_{\rm{Pb-Pb}}/T_{\rm{S-S}}\approx $ 15.4, which is indicative of
the origin of these photons basically from a collision mechanism. (Note that
the time-like partons are generated only if there is a collision.)
The slight deviation from this scaling seen at lower
$ p_T $ results in a $ \approx $ 20\% increase in the 
integrated yield at central rapidities. This is a good measure of the multiple
scatterings in the PCM.  In fact we have found that the number of hard
scatterings in the $ Pb+Pb $ system  is $ \approx $ 17 times more
than that for the $ S+S $ system  which also essentially determines the
ratio of the number of the photons produced in the two cases.
We also note that the inverse slope of the $ p_T $ distribution
decreases at larger rapidities, which is suggestive of the fact that
the ``hottest'' partonic system is formed at central rapidities.

\begin{figure}[b!]
\centerline{ \epsfig{file=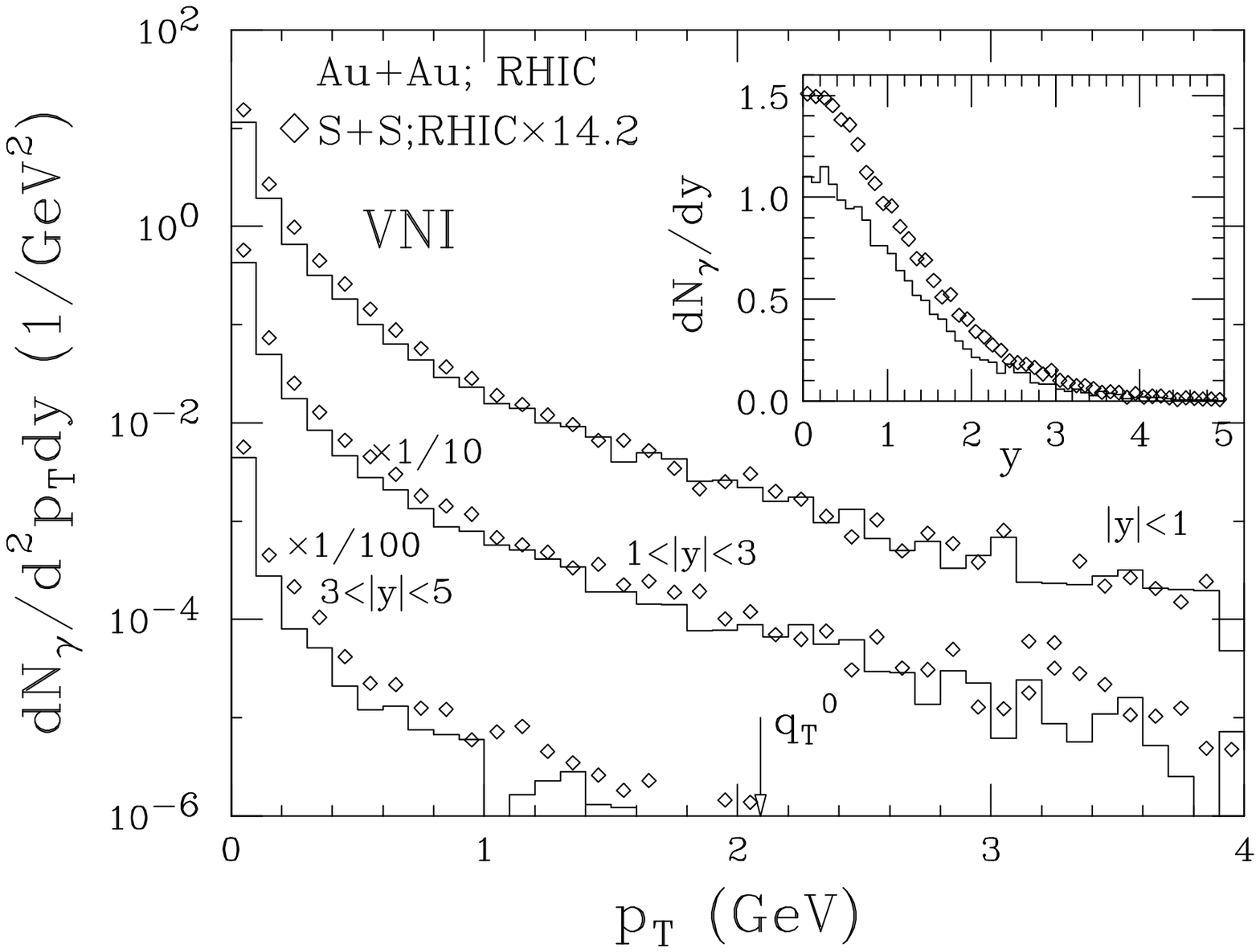,height=5.5in,width=5.5in}}
\vspace{10pt}
\begin{caption}
The radiation of single photons from S+S
and Au+Au collisions in different rapidity bins at RHIC energy,
\protect{$\sqrt{s}$} = 200 GeV/nucleon.
The inset shows the rapidity distribution of the radiated photons.
The results for the S+S collisions have been scaled by the
ratio \protect{$T_{Au-Au}/T_{S-S}\approx$} 14.2 for central collisions.
\protect{$\mu_0$}
for the quark branching \protect{$q\rightarrow\,q\gamma$}
is taken as 0.01 GeV.
\end{caption}
\label{fig3}
\end{figure}
In Fig.~3 we have plotted our results for $ S+S $ and $ Au+Au $ systems
at RHIC energies in the same fashion as Fig.~2 above. We see that the
inverse slope of the $ p_T $ distribution is now larger and 
drops only marginally at larger  rapidities,
indicating that the  partonic system is now ``hotter'' and
spread over a larger range of rapidity.
Even though the $ p_T $ distribution of the photons is seen to
roughly scale with the ratio of the nuclear overlap functions
for central collisions $ T_{\rm{Au-Au}}/T_{\rm{S-S}}\approx $ 14.2,
the integrated yield of photons for the $ Au+Au $ is seen to be only about 
12 times  that for the $ S+S $ system at the RHIC energies. We have
again checked that the number of hard scatterings for the
$ Au+Au $ system is also only about 12 times that for the $ S+S $
system. (Again note that we have switched off soft-scatterings
completely, and the hard scatterings are permitted
only if the $ p_T $ is more than $ p_T^0 $ which is taken to be larger 
at higher energies; see~\cite{p0}.) 

This contrasting behaviour at SPS and RHIC energies seen in our
work has a very interesting physical origin. At the SPS
energies, the partonic system begins to get dense 
and multiple scatterings increase; specially for heavier
colliding nuclei. At RHIC energies, the partonic system gets
quite dense, and then the Landau Pomeranchuk effect starts
playing an important role. We have implemented this in the 
PCM semi-phenomenologically; by inhibiting a new scattering
of partons till the passage of their
formation time after a given scattering.
In a separate publication we have demonstrated that these competitive
mechanisms can be seen at work by comparing results for zero
impact parameter for different colliding nuclei.

\section*{Discussions and summary}

Before concluding, we would like to make some other observations.

Firstly, recall such branchings of the partons produced in hard collisions
correspond to a next-to-leading-order correction in $ \alpha_s $.
These are known to be  considerably enhanced for collinear
emissions. The parton shower mechanism incorporated in the PCM
amounts to including these enhanced contributions to all orders,
instead of including all the terms for a given order~\cite{rkellis}.

It may also be added that the first-order corrections to the
Compton and annihilation processes in the plasma have
been studied by a number of authors~\cite{pradip}; however
in the plasma the $ Q^2\,\approx\,(2T)^2 $, thus their contribution is limited
to very low $ p_T $. $ Q^2 $ is obviously much larger in the early
hard scatterings,  and thus the radiations from the emerging
partons are much more intense and also populate higher transverse momenta,
as seen in the present work.

The large yield of photons from the branching of energetic quarks
preceding the formation of dense partonic matter opens an interesting
possibility to look for a similar contribution to dilepton (virtual
photon) production in such collisions.
In fact a large yield of low-mass dileptons
was reported from $ q\,\rightarrow\,q\,\ell\bar{\ell} $ processes in
 PCM~\cite{pcm6} calculations at RHIC energies.
It is quite likely that this process also
makes a  substantial contribution to the ``excess'' low mass dileptons
observed in sulfur and lead induced collisions studied 
at SPS energies. 

\begin{figure}[b!]
\vspace{60pt}
\centerline{ \epsfig{file=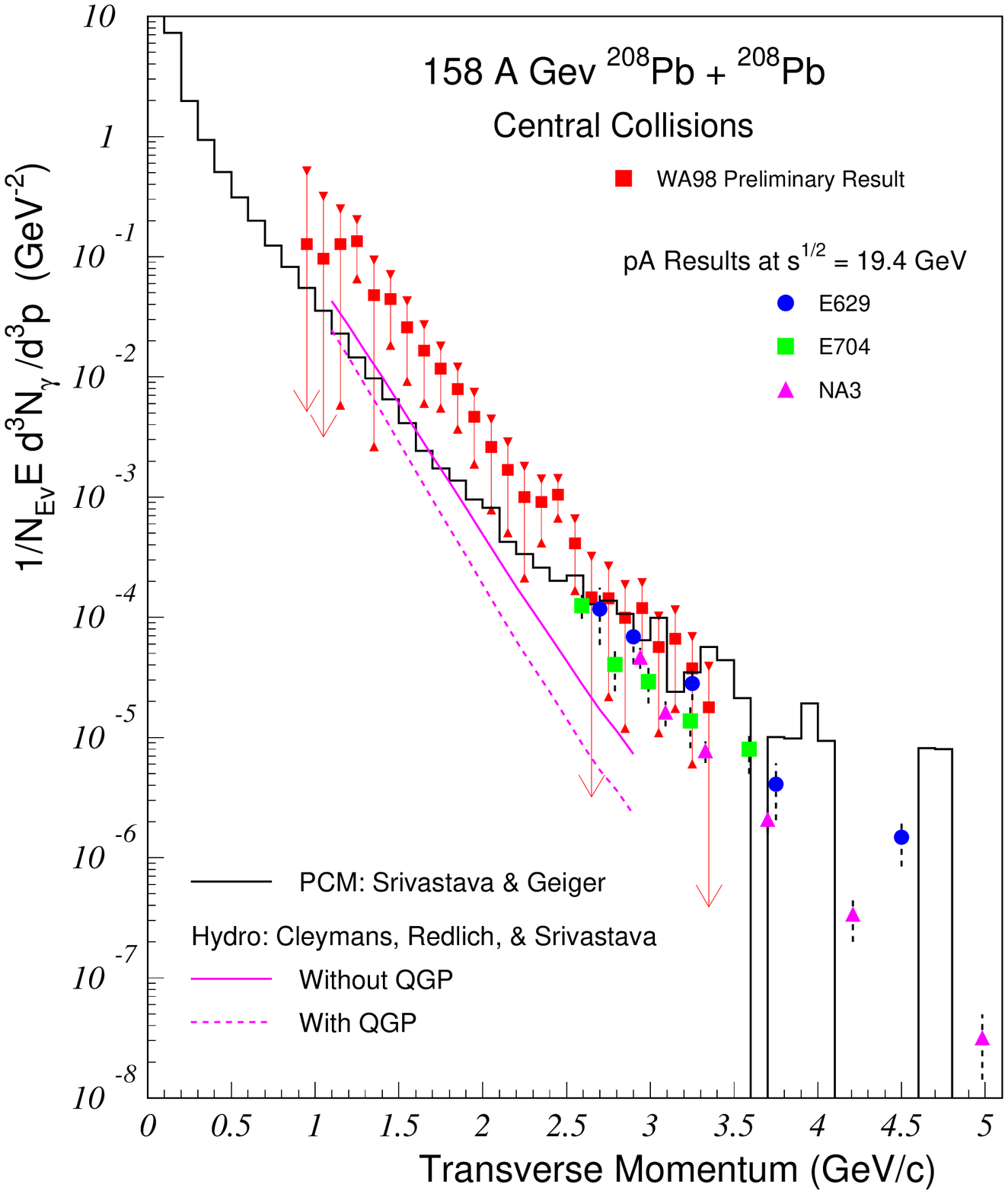,height=3.5in,width=3.5in}}
\vspace{80pt}
\begin{caption}
A comparison of the radiation of single photons from 
Pb+Pb collision at SPS energies with various predictions.
The (preliminary) data are from the WA98 collaboration. The histogram
shows the radiations from the partonic matter (only) evaluated
in this work. The solid curve gives the predictions of Cleymans,
Redlich, and Srivastava~\protect{\cite{crs}} when a hot hadronic matter is
assumed to be formed at 1 fm/\protect{$c$} evaluated within a hydrodynamic
model. The dashed curve gives their prediction when a thermalized
and chemically equilibrated quark gluon plasma is assumed to be
formed at 1 fm/\protect{$c$}. While the PCM predictions do not account
for the hadronic contributions to single photons, the 
hydrodynamic predictions do not account for the contributions
before the time \protect{$\tau=$} 1 fm/\protect{$c$}.
\end{caption}
\label{fig4}
\end{figure}

Recall again that  we have only included the contribution of
photons from the partonic interactions in this work. It is
quite likely that the hadrons produced at the end will also
interact and produce photons which has been extensively
studied in recent times (see, e.g. Ref.~\cite{crs}). A comparison of
typical results from, say,  Ref.~\cite{crs} with the present work
shows that at the SPS energies the emission from the early
hard partonic scatterings is of the same order as 
the photon production  from later hadronic reactions, for $ p_T\leq $ 2--3 GeV,
and dominates considerably over the same at higher transverse momenta. 
A comparison of our predictions (Fig.4) with the preliminary results reported
by the WA98 collaboration~\cite{WA98} in fact clearly demonstrates that
the pre-equilibrium contributions evaluated in the present
work will play a very important role in providing a proper 
description to the single photon data at lager $p_T$ from such collisions.

We conclude that the formation of a hot and dense partonic
system in relativistic heavy ion collisions may be preceded
by a strong flash of photons following the early hard scatterings.
Their yield will, among several other intersting aspects, also
throw light on the extent of multiple scattering encountered
in these collisions.

\bigskip

\section*{ACKNOWLEDGEMENTS}
Most of this work was done when I visited Klaus at the Brookhaven
National Laboratory during December 15, 1997 to March 15, 1998. Little
did I know that I would not see Klaus again. I was in e-mail contact
with him till a day before he departed. I wondered what should I 
write, having done several things and having planned several things in
collaboration with him and yet not being able to attend the workshop
in his memory. The choice was not easy. This work appeared in
print~\cite{pcm7} in the September 1998 issue of the Physical Review C;
with Klaus going away in a `flash' just when these issues were perhaps
being mailed. Yes Klaus;

\begin{verse}
The world was listening\\
With so much attention,\\
Alas! you dozed off\\
While telling your tales..\\
\end{verse}

Boss!, I will always miss you, remember you, and endeavour to complete
all that we planned.

This work was supported in part by the D.O.E under contract no.
DE-AC02-76H00016.

\end{document}